\documentclass[aps,pre,twocolumn,showpacs,groupedaddress]{revtex4}  
\usepackage{graphicx}  
\usepackage{dcolumn}   
\usepackage{bm}        
\usepackage{amssymb}   
\usepackage{float}
\usepackage[normalem]{ulem}
\usepackage{enumerate}

\begin{document}
\hspace{5.2in}
\title{Different kinds of chimera death states in nonlocally coupled oscillators}
\author{K. Premalatha$^{1}$, V. K. Chandrasekar$^{2}$, M. Senthilvelan$^{1}$, M. Lakshmanan$^{1}$}
\address{$^1$Centre for Nonlinear Dynamics, School of Physics, Bharathidasan University, Tiruchirappalli - 620 024, Tamilnadu, India.\\
$^2$Centre for Nonlinear Science \& Engineering, School of Electrical \& Electronics Engineering, SASTRA University, Thanjavur -613 401,Tamilnadu, India.}
\begin{abstract}
We investigate the significance of nonisochronicity parameter in a network of nonlocally coupled Stuart-Landau oscillators with symmetry breaking form.  We observe that the presence of nonisochronicity parameter leads to structural changes in the chimera death region while varying the strength of the interaction.  This gives rise to the existence of different types of chimera death states such as multi-chimera death state, type-I periodic chimera death (PCD) state and, type-II periodic chimera death (PCD) state.  We also find that the number of periodic domains in both the types of PCD states exponentially decreases with an increase of coupling range and obeys a power law under nonlocal coupling.  Additionally, we also analyze the structural changes of chimera death states by  reducing the system of dynamical equations to a  phase model through the phase reduction.  We also briefly study the role of nonisochronicity parameter on chimera states, where the existence of multi-chimera state with respect to the coupling range is pointed out.  Moreover, we also analyze the robustness of the chimera death state to perturbations in the natural frequencies of the oscillators.
\end{abstract}
\maketitle
\section{Introduction}
\par Coupled nonlinear oscillators can exhibit a variety of interesting dynamical phenomena such as synchronization, chaos, clustering, chimera states, complex patterns and so on \cite{1}.  These phenomena have been extensively studied in various fields of physics, biology, and the other branches of science and technology \cite {2}.  Beside these complex behaviors, oscillation death (OD) can also occur which is a phenomenon in which the oscillations decrease and ultimately converge to zero.  OD is characterized by a stabilization of two stable inhomogenous steady states under different coupling schemes such as mean field diffusion coupling \cite{3,4,5}, dynamic coupling \cite {6} and, conjugate coupling \cite{7} in identical oscillators, delay coupled identical oscillators \cite{7a}, and dynamic environment coupling \cite{8} in identical or mismatched oscillators.
\par On the other hand, nonlocal coupling topology in coupled networks leads to the phenomenon of chimera states\cite {11,12,13,14,15,16,17,18,19,20,21,22,25,26,22a}.  In a chimera state, regions with spatial coherence coexist with regions of spatial incoherence in nonlocally coupled identical oscillators.  Recently, such coexistence behavior of oscillators has led to considerable attention concerning the importance of chimera states in real life applications.  These observations of chimera states have helped to explain various phenomena that occur in practice, including uni-hemispheric sleep \cite{27}, power distribution in networks \cite{28}, and bump states in neural networks \cite{29} and related systems \cite{30,31,32,33,34,35,36}.  
\par Interestingly, the interplay of nonlocality with symmetry breaking in the coupling gives rise to another emergent phenomenon, namely chimera death state, which was observed first by Anna Zakharova et al. \cite{37}.  In this case the tendency of symmetry breaking in the system in inducing oscillation death(OD) and the tendency of nonlocal coupling in inducing the chimera states coalesce resulting in a new state called chimera death.   In the chimera death state,  the oscillators in the network partition into two coexisting domains, where in one domain neighboring nodes occupy the same branch of the inhomogeneous steady state (spatially coherent OD) while in the other domain neighboring nodes are randomly distributed among the different branches of inhomogeneous steady state (spatially incoherent OD).  While tuning the coupling parameters multi-chimera death states can also occur as multi-coherent and incoherent distributions of inhomogeneous steady states (here by inhomogeneous steady states we mean the two branches of a stable steady state, one follows the upper branch while other follows the lower branch).  Anna Zakharova et al. have explored this phenomenon in nonlocally coupled Stuart-Landau oscillators in the absence of amplitude dependent frequency parameter (nonisochronicity parameter).  Recently the same phenomenon was also observed in globally coupled Stuart-Landau oscillators in the presence of nonisochronicity parameter \cite{37a} as well as in the absence of nonisochronicity parameter \cite{37b}.  
\par In the present work, we address the question as to how the structure of chimera/multi-chimera death state is affected by the nonisochronicity parameter and also by the coupling range in the case of nonlocal coupling.  To illustrate the above results, we again consider an one dimensional array of Stuart-Landau (SL) oscillators that are coupled nonlocally with symmetry breaking form of interaction.  We observe the formation of different kinds of chimera death states only in the presence of nonisochronicity parameter as a function of coupling strength.  Interestingly, we find different types of chimera death states, including multi-chimera death state, type-I periodic chimera death (PCD) state, and type-II periodic chimera death (PCD) state.  Note that in a multi-chimera state, more than one group of coherent and incoherent oscillation death (OD) states occur but they do not occur in a periodic manner. On the other hand, in the case of PCD states the distribution of OD states exist periodically.  Moreover, the number of periodic domains ($n_0$) in the periodic chimera death region obeys a power law relation with the coupling range.  Additionally, we analyze the structural changes of chimera death states by reducing the system of dynamical equations to a phase model through the phase reduction.  On the other hand, we also identify that the chimeras which appear in the oscillation region do not follow any relation with the coupling range.  We also examine the nature of the chimera death states as a function of the distribution of natural frequencies of the oscillators.    
 \par This paper is organized as follows.  In section II, we introduce the model of nonlocally coupled Stuart-Landau oscillators that we have considered for our simulation and present the results obtained from the analysis of different chimera death states under the influence of the nonisochronicity parameter. Next we have analyzed the  occurrence of the chimera death state for different coupling ranges in section III.  In section IV, we study the dynamics of the system by using a phase reduction.  In section V, the dynamics of the oscillatory states in the presence of nonisochronicity parameter is investigated in some detail.  We also analyzed the robustness of chimera death states with perturbation of natural frequencies in section VI.  We summarize our findings in section VII.  In appendix-A, we indicate the distinguishing features of the multi-chimera, type-I and type-II periodic chimera death states, while in appendix-B the basic features of the characteristic measures, namely the strength of incoherence  and discontinuity measure are briefly explained.
\section{CHIMERA DEATH STATES IN COUPLED STUART-LANDAU OSCILLATORS UNDER NONLOCAL COUPLING WITH SYMMETRY BREAKING}  
\subsection{MODEL} 
\par We consider a ring of nonlocally coupled identical Stuart-Landau oscillators with symmetry breaking in the coupling, whose dynamics can be represented by the following set of equations,
\begin{equation}
\dot{z_j}=(1+i \omega)z_j-(1- ic)|z_j|^2z_j \nonumber\\+\frac{\varepsilon}{2P} \sum_{k=j-P}^{j+P} (Re[z_k]-Re [z_j]),
\label{n}
\end{equation}
where $z_j=x_j+iy_j$, $j=1,2,3,...N$.  Here $\omega$ is the natural frequency of the oscillators, $c$ is the nonisochronicity parameter, and $N$ is the total number of oscillators.  The nonlocal coupling in the system is controlled by the coupling strength ($\varepsilon$) and the coupling range or radius ($r=\frac{P}{N}$), where $P$ corresponds to the number of nearest neighbors in both the directions.  Here, we have introduced the coupling only in the real parts of the complex amplitude, and so this coupling introduces a symmetry breaking in the system. 
\par  In our simulations, we choose the number of oscillators $N$ to be equal to 500, $\omega=3$ and in order to solve the Eq. (\ref{n}) numerically, we use the fourth order Runge-Kutta method with a time step 0.01 and the initial state of the oscillators $(x_j,y_j)$ are independently distributed with uniform random values between -$1$ and $+1$.
\subsection{STUDY OF STRUCTURAL CHANGES IN THE CHIMERA DEATH STATES}	
\par In this section, we study the consequences of frequency dependent amplitude (nonisochronicity) parameter ($c$) on the formation of chimera death states.  A chimera death state represents spatial coexistence of coherent and incoherent oscillation death states.  A coherent OD represents the population of neighboring oscillators in the same branch of inhomogeneous steady state while incoherent oscillation death corresponds to the case when the sequence of populated branches of inhomogeneous steady states for the neighboring elements is random.  In our studies, we observe that nonisochronicity is the key for the existence of a variety of a dynamical states.  Generally,  the nonisochronicity plays a crucial role for the onset of complex behavior in an ensemble of identical oscillators.  To explore the role of c on chimera death states, we begin by choosing the coupling range as $r=0.4$ in the system (\ref{n}).
\par
\begin{figure}[ht!]
\begin{center}
\includegraphics[width=1.0\linewidth]{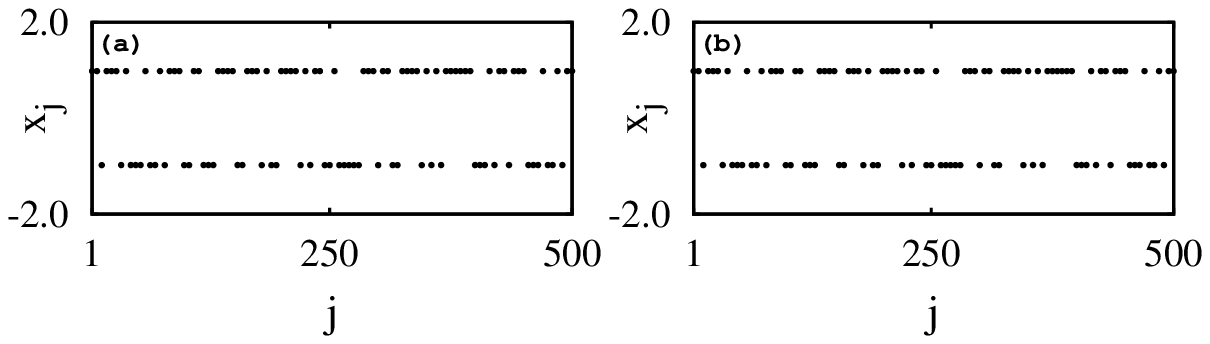}
\end{center}
\vspace{-0.8cm}
\caption{ Multi-chimera death state for $r=0.4$: (a) $\varepsilon=12$ ($\frac{\varepsilon}{2P}=0.03$), (b) $\varepsilon=32$ ($\frac{\varepsilon}{2P}=0.08$) in the absence of $c$.}
\label{non}
\end{figure}

  It is known from Ref. \cite{37} that the chimera death states in the absence of nonisochronicity parameter ($c=0$) appear as number of clusters in the coherent domains and there is no incoherent domain which appears in between the coherent clusters.  These states are designated as multi-cluster chimera death states.  Note that the initial conditions chosen in ref. \cite{37} are such that the oscillators at the edges are randomly distributed (in the interval [0,1] for the upper group and [-1,0] for the lower group) while those in between are distributed uniformly (with half of them in the upper branch and the remaining in the lower branch) in either of the inhomogeneous steady states.  However in the present study, we observe the multi-coherent and multi-incoherent OD states for the initial conditions randomly distributed in the entire range between $-1$ and $+1$.  The resultant state is termed as multi-chimera death state.  Here in this steady state all the oscillators in the array are distributed uniformly either in the lower branch or in the upper branch (Fig. \ref{non}(a)) and phases of the oscillators are distributed with the difference $0$ or $\pi$.  By increasing the coupling strength, there occurs no change in the distribution of the inhomogeneous steady states.  Hence strengthening the coupling interaction does not affect the distribution of the multi-chimera death state and it is stable under the absence of the nonisochronicity parameter $c$ as illustrated in Fig. \ref{non}(b).  We now analyze the question of how the influence of nonisochronicity parameter leads to the onset of different types of chimera death states.  For this purpose we fix the value of $c$ as $c=4$ throughout this section (for the reason that the formation of different chimera death states occurs only for a sufficiently large value of $c$) for the entire coupling range. 
\begin{figure*}[ht!]
\begin{center}
\includegraphics[width=1.0\linewidth]{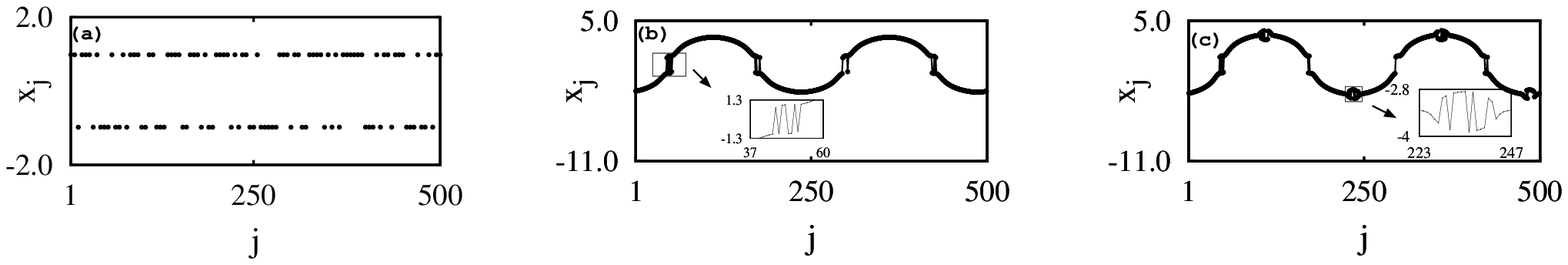}\\
\vspace{-1.0cm}
\includegraphics[width=1.0\linewidth]{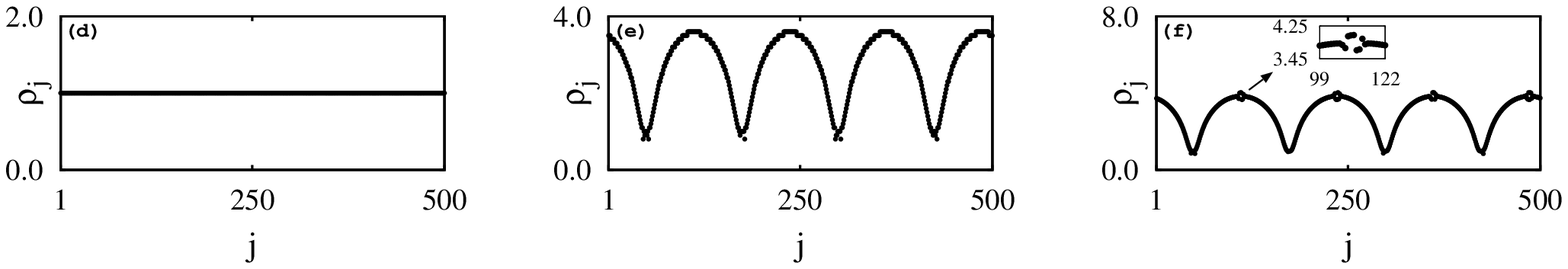}\\
\vspace{-1.0cm}
\includegraphics[width=1.0\linewidth]{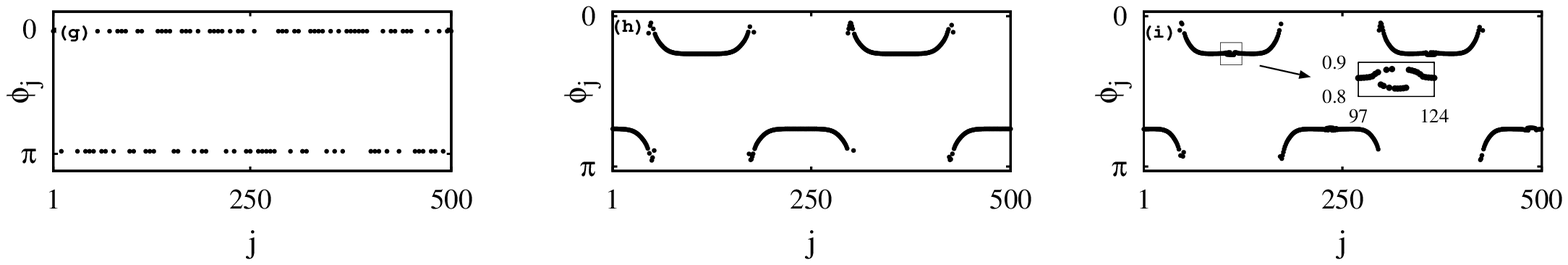}
\end{center}
\vspace{-0.6cm}
\caption{Different kinds of chimera death states, their corresponding amplitudes $\rho_j=\sqrt{x_j^2+y_j^2}$ and the associated phases $\phi_j=\arctan(y_j/x_j)$ for various states with $c=4$ and $r=0.4$: (a),(d),(g) multi-chimera death state for $\varepsilon=22.4$ ($\frac{\varepsilon}{2P}=0.056$), (b),(e),(h) type-I periodic chimera death state for $\varepsilon=23.2$ ($\frac{\varepsilon}{2P}=0.058$) and (c),(f),(i)  type-II periodic chimera death state for $\varepsilon=24$ ($\frac{\varepsilon}{2P}=0.06$).}
\label{fg2}
\end{figure*}  
\begin{figure*}[ht!]
\begin{center}
\includegraphics[width=1.0\linewidth]{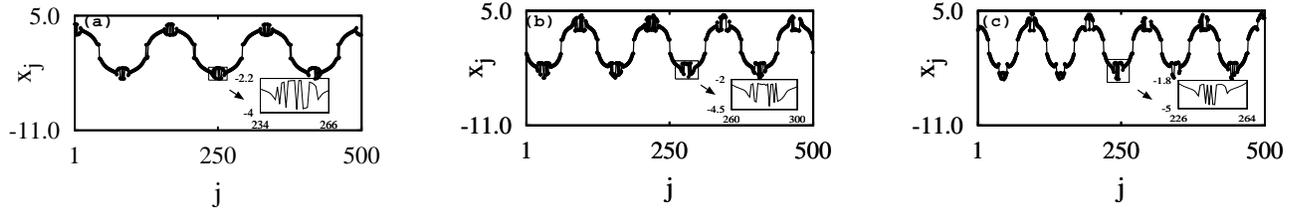}
\end{center}
\vspace{-0.8cm}
\caption{Type-II PCD with $c=4$ for different values of coupling range: (a) $r=0.25$ and $\varepsilon=18$ ($\frac{\varepsilon}{2P}=0.072$), (e) $r=0.2$ and $\varepsilon=22$ ($\frac{\varepsilon}{2P}=0.11$), (f) $r=0.15$ and $\varepsilon=26$ ($\frac{\varepsilon}{2P}=0.173$).}
\label{fg1}
\end{figure*}

\begin{figure}[ht!]
\begin{center}
\includegraphics[width=0.75\linewidth]{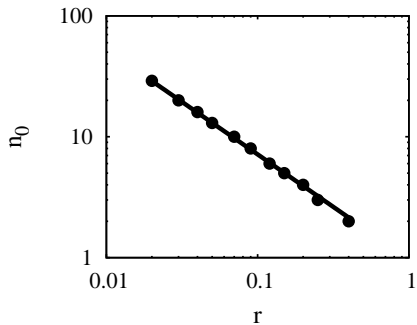}
\end{center}
\caption{Number of periodic domains in the PCD states $n_0$ as a function of the coupling range $r~(=\frac{P}{N})$ in logarithmic scale.  Dot represents the numerical data and the corresponding best fit is represented by the (black) curve.}
\label{num}
\end{figure} 
\begin{figure}[ht!]
\begin{center}
\includegraphics[width=1.05\linewidth]{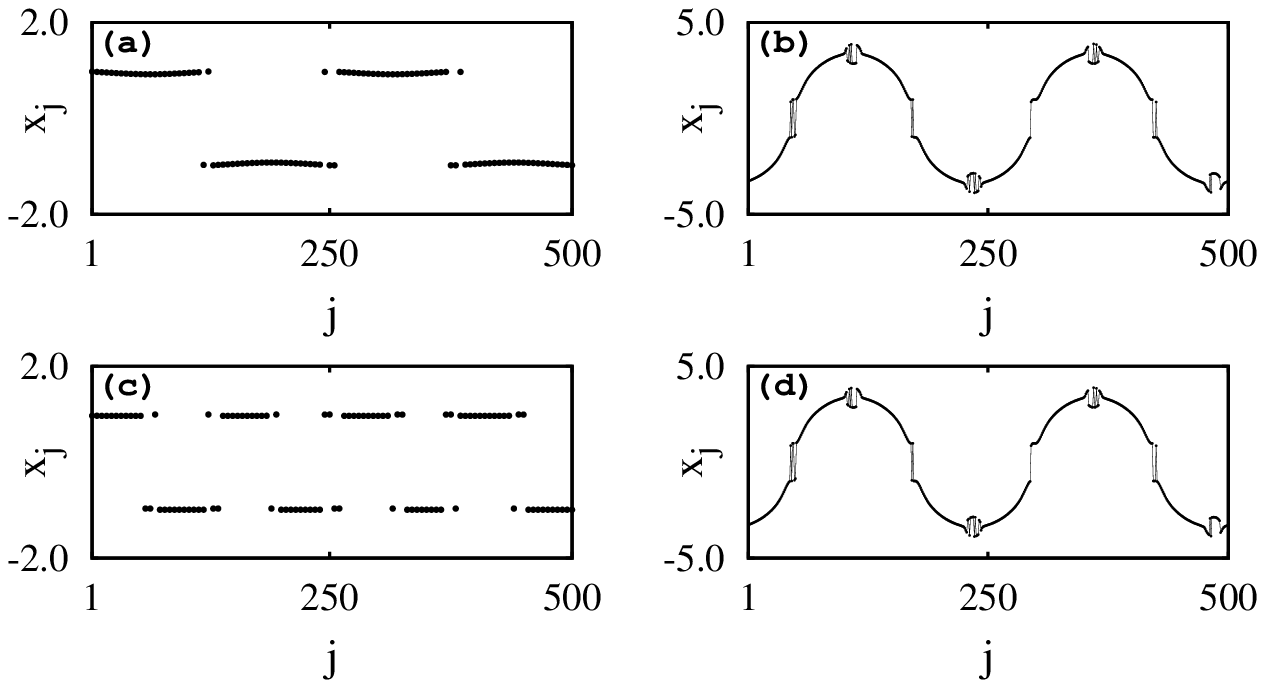}
\end{center}
\caption{(a) 2-chimera death state for $\varepsilon=22.4$ ($\frac{\varepsilon}{2P}=0.056$), (b) type-II 2-PCD state for $\varepsilon=24$ ($\frac{\varepsilon}{2P}=0.06$) corresponding to a distribution of 2-clusters in the initial condition, (c) 5-chimera death state for $\varepsilon=22.4$ ($\frac{\varepsilon}{2P}=0.056$) and (d) type-II 2-PCD state $\varepsilon=24$ ($\frac{\varepsilon}{2P}=0.06$) corresponding to a distribution of 5-clusters in the initial condition with $r=0.4, c=4$ and the other parameter values are same as mentioned above.}
\label{initial}
\end{figure}
\begin{figure*}[ht!]
\begin{center}
\includegraphics[width=0.8\linewidth]{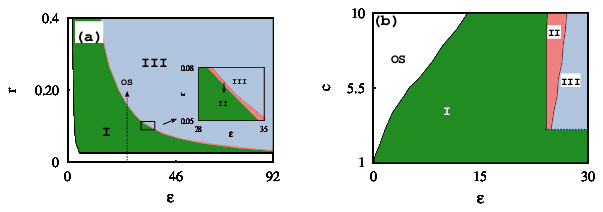}
\end{center}
\caption{(Color online) (a) Phase diagram in ($\varepsilon, r$) parametric space with $N=500$ oscillators for fixed $c=4$.  (b) Phase diagram in ($\varepsilon, c$) parametric space for fixed $r=0.4$.  (In Fig. \ref{fg4}(b)  $\frac{\varepsilon}{2P}$ varies between $0$ and $0.075$).  OS represents the oscillatory state, region-I represents the multi-chimera death state, region-II represents the type-I PCD state, region-III represents the type-II PCD state.  Note that the OS state consists of desynchronized, synchronized, amplitude chimera and frequency chimera states (which are not explicitly shown here) and are depicted explicitly in Fig. \ref{fg5}(b) below and discussed in detail in section V.}
\label{fg4}
\end{figure*} 

\par  In Fig. \ref{fg2}(a-c) we have presented the multi-chimera death states for different values of coupling strength.  First, we can observe in Fig. \ref{fg2}(a) that for the value $\varepsilon =22.4$ ($\frac{\varepsilon}{2P}=0.056$), a uniformly distributed set of inhomogeneous steady states (where the disorder induced by the nonisochronicity parameter is balanced by the effect of coupling strength for weak coupling interaction) occurs which may be identified as a multi-chimera death state.  Here the amplitudes ($\rho_j=\sqrt{x_j^2+y_j^2}$) of all the oscillators are practically the same and hence there is effectively no variation in the amplitudes  (Fig. \ref{fg2}(d)) and phases ($\displaystyle \phi_j=\arctan(y_j/x_j)$) of the oscillators are distributed with the difference $`0'$ and $`\pi'$ (Fig. \ref{fg2}(g)).  Another feature to be noted is that the total number of oscillators occupying the two distinct groups of homogeneous steady states is conserved which implies that $N_U=N_L$ ($N_U$ represents the total number of oscillators in the upper branch and $N_L$ represents the total number of oscillators in the lower branch) with $\frac{1}{N}\sum_{j=1}^N z_j=0$.  Next on increasing the value of the coupling strength $\varepsilon$, as may be seen in Figs. \ref{fg2} (b), (e), (h) and \ref{fg2} (c), (f), (i), we observe that an increase in disorder in the distribution of inhomogeneous steady states occurs due to the role played by the variation in the amplitudes and phases of the oscillators. This feature gives rise to the disappearance or a decrease in the number of incoherent OD regions and we can observe the existence of two spatially periodic domains of coherent and incoherent OD state (along the array) in each of the branches.  In contrast to multi-chimera death state, we can observe in the present case deviations in the amplitudes (and phases of the oscillators which will be discussed later in this section) of the oscillators.  Further, here, the incoherent steady states occupy the edges of the coherent domains.  So we call this state as a type-I 2-periodic chimera death (2-PCD) state which is shown in Fig. \ref{fg2}(b) for $\varepsilon=23.2$ ($\frac{\varepsilon}{2P}=0.058$).  On increasing the coupling strength further, one finds a further increase in disorder in the amplitudes of the oscillators which initiates the formation of incoherent domains within each coherent domain also.  The incoherent domains which are located at the edges of the coherent domains remain unchanged.  This forms a new kind of steady state, namely type-II 2-PCD state as shown in Fig. \ref{fg2}(c) for the coupling strength $\varepsilon=24$ ($\frac{\varepsilon}{2P}=0.06$).  The appearance of incoherent domains located in the middle of the coherent domains is clearly seen in the inset of Fig. \ref{fg2}(c).  In the case of this new kind of chimera death state also the total number of oscillators are equally split into two branches of the inhomogeneous steady states ($N_U=N_L$) with $\frac{1}{N}\sum_{j=1}^N z_j=0$.  Thus we conclude that the presence of nonisochronicity parameter leads to an increased nonuniformity and amplitude variations in the chimera death states and causes the existence of different chimera death states, namely type-I and type-II 2-PCD states.
\par  In order to validate the identification of different chimera death states, we analyze the amplitudes of each of the oscillators  by using the expression for $\displaystyle \rho_j$, which are shown in Figs. \ref{fg2}(d-f) (with the same parameter values as used in Figs. \ref{fg2}(a-c)).   Fig. \ref{fg2}(d) shows that the oscillators are having the same amplitude for weak coupling interaction corresponding to a multi-chimera death state when $\varepsilon<\varepsilon_c$.  Here $\varepsilon_c=22.8~ (\frac{\varepsilon}{2P}=0.057)$ denotes the critical value of the coupling strength at which the transition from the multi-chimera death state (where the amplitude variation is negligible) to type-I PCD state (where the amplitude variation is appreciable) occurs for the coupling range $r=0.4$.  Increasing  the strength of the coupling interaction gives rise to amplitude variations due to the presence of the nonisochronicity parameter $c$ leading to a type-I 2-PCD state.  The amplitudes of the oscillators in the upper and lower branches are periodically modulated in space which is illustrated in Fig. \ref{fg2}(e).  It can also be seen that in Fig. \ref{fg2}(f), the amplitude deviations get further increased in the strong coupling limit ($\varepsilon>\varepsilon_c$) which results in a type-II 2-PCD state where all the oscillators in the periodic domain do not have the same amplitude and we can observe the coherent and incoherent behaviors in the amplitude distributions (the amplitude variations are also explained with phase portraits in appendix-A).
\par To give a better understanding about the phase dynamics of different chimera death states, we find the phase of each of the oscillators by using the expression for $\displaystyle \phi_j$, where $j=1,2,3, ...N$.  These are shown in Figs. \ref{fg2}(g-i).  In the multi-chimera death region the phases of the oscillators are distributed with the difference $`0'$ and $`\pi'$ (Fig. \ref{fg2}(g)).  An increase in $c$ leads to deviations not only in the amplitudes but also in the phases of the oscillators where the phase differences now lie between $`0'$ and $`\pi'$.  In the case of type-I PCD state, the phases are nearly the same at the center of the coherent domains and there occurs a random distribution near their edges which is shown in Fig. \ref{fg2}(h).  It can also be seen that in Fig. \ref{fg2}(i), we observe incoherent distributions of phases in the middle as well as near edges of the coherent domains for the case of the type-II PCD state.  Thus nonisochronicity causes deviations in the amplitudes as well as in the phases of the oscillators.  We can observe the deformation of steady states for the choice of the initial condition considered in Ref. \cite{od} in the absence of nonisochronicity parameter.  However we can observe the type-I and type-II periodic domains of chimera death states only in the presence of nonisochronicity parameter.  In this section we have studied the existence of different kinds of chimera death states for fixed value of $c$ by varying $\varepsilon$.  We can observe the different transition scenarios by varying c itself for fixed but large values of $\varepsilon$ also.  More details are given in the following sections.

\section{Occurrence of chimera death state with respect to coupling range and scaling laws}
\par   Next, we analyze the question of how the structure of chimera death states change with respect to the coupling range ($r$) {in the presence of $c$.  Interestingly, we can observe that the number of periodic coherent OD and incoherent OD domains increases in the strong coupling limit when $\varepsilon>\varepsilon_c$ (regions corresponding to type-I and type-II PCDs) while decreasing the coupling range $(r)$.  However, there occurs no change in the multi-chimera death states in the weak coupling limit, $\varepsilon<\varepsilon_c$.  We illustrate these facts specifically in type-II PCD (for the reason that we can observe the same number of periodic domains in both the type-I and type-II PCDs) with different values of coupling ranges, namely $r=0.25, 0.20, 0.15$.  In the case of $r=0.25$, we can observe three periodic domains of chimera death state in each of the branches as shown in Fig. \ref{fg1}(a).  Decreasing the coupling range to $r=0.2$ (in Fig. \ref{fg1}(b)) and $r=0.15$ (in Fig. \ref{fg1}(c)), we can observe the increase of periodic domains that is type-II 4-PCD and type-II 5-PCD states, respectively.  Thus decreasing the coupling range leads to the creation of periodic coherent and incoherent OD regions in the strong coupling limit.  However, we find that there is no change in the  distribution of multi-chimera death states which is not shown here.  
\par Fig. \ref{num} shows the log-log plot of the number of periodic domains ($n_0$) against the coupling range ($r=P/N$).  Existence of number of periodic domains follow a power law relation $n_0=a r^b$ against the coupling range and the best curve fit is obtained for the values $a= 0.929446$ and $b= -0.87964 $.  It can be seen in Fig. \ref{num} that the number of periodic domains exponentially decreases with an increase of the coupling range ($r$).  Moreover, type-I and type-II PCD states persist their structures to different forms of initial states of the oscillators while the chimera death state in the weak coupling region depends on or takes the form as the initial states of the oscillators.  For example in our study we chose random initial conditions between -1 and +1 for $r=0.4$ and we get the type-I 2-PCD and type-II 2-PCD as shown in Figs. \ref{fg2} (b), (c).  To confirm the robustness in formation of the PCD states for different clusters of size $n$ between $-1$ to $+1$ as initial conditions, we choose the specific case of $0< z_{(n-1)m+1}, z_{(n-1)m+2},...z_{(n-\frac{1}{2})m}<1$, $-1<z_{(n-\frac{1}{2})m+1}, z_{(n-\frac{1}{2})m+2},...z_{nm}<0$, where $ m=\frac{N}{n}$, $N$ is the total number of oscillators, $m$ is the cluster size and $n$ is the number of clusters.  It can be seen in Figs. \ref{initial} (a, c) that the chimera death state in the weak coupling region depends on or takes the form as the initial states of the oscillators.  As an example, for the 2-clusters of distribution ($n=2$) in initial conditions, one can get the 2-chimera death state (Fig. \ref{initial}(a)) while a 5-cluster distribution of initial condition ($n=5$) leads to the existence of 5-chimera death state for $\varepsilon=22.4$ ($\frac{\varepsilon}{2P}=0.056$) in Fig. \ref{initial}(c).  On the other hand, we do not observe a change in the distribution, that is the number of periodic domains in the PCD with respect to the form of the initial conditions [figs. \ref{initial} (b,d)] for $\varepsilon=24$ ($\frac{\varepsilon}{2P}=0.06$).  Hence we have confirmed that PCD states persist their structure for different forms of initial conditions.  Thus, coupling range plays a crucial role over the existence of the number of periodic domains.  Other parameters such as natural frequencies of the oscillators, nonisochronicity parameter and also the form of the initial condition do not influence the number of periodic domains of the chimera death states.  

\par  To give a better insight about the chimera death states, we have plotted the two phase diagram in the ($\varepsilon,r$) parametric space for $c=4$ in Fig. \ref{fg4}(a), using the same numerical protocol mentioned in Sec. II. A. ( That is, $N=500$, $\omega=3$ using fourth order Runge-Kutta method with a time step 0.01 and the notion of strength of incoherence as explained in appendix-B.  The same procedure is also used in Fig. \ref{fg4}(b) and Fig. \ref{fg5} in the following).  For very low coupling range $r<0.02$, (nonlocal coupling), we can observe only oscillatory states and no chimera death states.  Similarly, for $r =0.5$, we cannot observe the presence of spatially periodic domains along the array in the chimera death state as a result of nonlocal coupling limit approaching the global limit.  Similarly, for very low coupling strength, only oscillatory states occur (more details about the characteristic nature of the dynamical states which occur in the oscillatory region are discussed in Fig. \ref{fg5}(b) and section V below).  In the remaining regions we identify multi-chimera death (region-I), type-I PCD (region-II), type-II PCD (region-III) in the ($\varepsilon, r$) phase portrait, Fig. \ref{fg4}(a).  In order to give a consolidated picture of different chimera death states, we also present a two parameter phase diagram in the ($\varepsilon,c$) parametric space for $r=0.4$ in Fig. \ref{fg4}(b).  Here we can observe that for a sufficiently small value of the nonisochronicity parameter ($c<3.0$) there occurs no change in the distribution of the chimera death states and we can observe only multi-chimera death states.  With an increase in the nonisochronicity parameter to significant values,  we can observe the change in the distribution in the multi-chimera death region and the onset of type-I PCD regions as well as type-II PCD regions.  From our investigations, we infer that the presence of the nonisochronicity parameter leads to new kinds of chimera death states.

\begin{figure}[ht!]
\begin{center}
\includegraphics[width=1.0\linewidth]{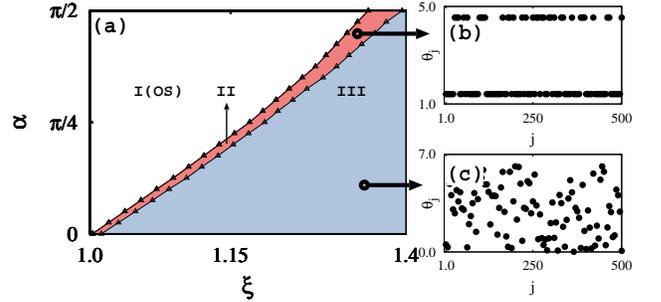}
\end{center}
\caption{(Color online) (a) Phase diagram of the nonlocally coupled system (\ref{nonlocalp}) in the parametric space ($\xi, \alpha$).  (b) and (c) represent the instantaneous phases of all the oscillators for $\xi=1.35$ which are marked by ($\circ$) for $\alpha=\frac{7\pi}{16}$ in the region-II (red/dark-grey) and for $\alpha=\frac{\pi}{8}$ in the region-III (grey), respectively.}
\label{ran11}
\end{figure} 
\begin{figure}[ht!]
\begin{center}
\includegraphics[width=1.0\linewidth]{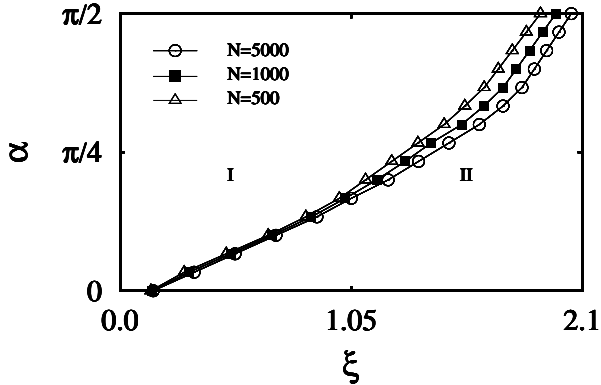}
\end{center}
\caption{(a) Phase diagram of the globally coupled system in the parametric space ($\xi, \alpha$) obtained from Eq. (\ref{t}).  From Eq. (\ref{t}), the solid line with ($\vartriangle$) corresponds to the boundary of the oscillatory state (region-I) and steady state (region-II) for $N=500$, the solid line with ($\blacksquare$) corresponds to the boundary for $N=1000$, the solid line with ($\circ$) represents the boundary for $N=5000$.}
\label{curve1}
\end{figure} 
\section{Reduction to a phase model}
\par In this section, we analyze the dynamics by reducing Eq. (\ref{n}) to a phase model by rewriting it as follows, 
{\small
\begin{eqnarray}
\dot{\rho_j}=(1-\rho_j^2)\rho_j+\frac{\varepsilon}{2P}\sum_{k=j-P}^{j+P}[\frac{\rho_k}{2}(\cos(\theta_k-\theta_j)+\cos(\theta_k+\theta_j))\nonumber\\
-\frac{\rho_j}{2}(1+\cos(2\theta_j))] 
\end{eqnarray}}
{\footnotesize
\begin{eqnarray} 
\rho_j\dot{\theta_j}=(\omega -c\rho_j^2)\rho_j+\frac{\varepsilon}{2P}\sum_{k=j-P}^{j+P}[\frac{\rho_k}{2}(\sin(\theta_k-\theta_j)-\sin(\theta_k+\theta_j))\nonumber\\
+\frac{\rho_j}{2}(\sin(2\theta_j))], j=1,2,...N, 
\end{eqnarray}}
where $z_j=\rho_j e^{i\theta_j}$.  Reduction to the phase equations is obtained by considering $\dot{\rho_j}\approx 0$, by assuming $\rho_j \approx$ constant for all $j$. Then Eq. (2) becomes
\begin{eqnarray}
\rho_j^2=1-\frac{\varepsilon}{4P}(2P+1)-\frac{\varepsilon}{4P}(2P+1) \cos(2\theta_j)\nonumber\\+ \frac{\varepsilon}{4P}\sum_{k=j-P}^{j+P}[(\cos(\theta_k-\theta_j)+\cos(\theta_k+\theta_j))].
\label{rho}
\end{eqnarray}
In such a case Eq. (3) is described by the following phase equation, after replacing $\rho_j^2$ by the right hand side of Eq. (\ref{rho}),
{\small
\begin{eqnarray}
\dot{\theta_j}=\omega -\tan \alpha+\frac{\varepsilon \tan \alpha }{4P}(2P+1)+\frac{\xi}{4P}(2P+1) \sin(2\theta_j-\alpha)\nonumber\\
+\frac{ \xi}{4P}\sum_{k=j-P}^{j+P}[(\sin(\theta_k-\theta_j-\alpha)-\sin(\theta_k+\theta_j+\alpha))], \qquad
\label{nonlocalp}
\end{eqnarray}}
where $c=\tan \alpha$ with $\alpha \geq \pi/2$ and $\xi=\frac{\varepsilon}{\cos \alpha}$.  Eq. (\ref{nonlocalp}) represents the phase oscillator model where the nonisochronicity is represented in the form of an asymmetry in the phase coupling parameter ($\alpha$).  
 Fig. \ref{ran11}(a) is plotted by solving Eq. (\ref{nonlocalp}) in the parametric space ($\varepsilon,\alpha$).  Instantaneous phases of the oscillators in the region-II (red/dark-grey) and region-III (grey) are illustrated in Figs. \ref{ran11}(b,c).  By distributing the initial conditions as $y_j^2=C-x_i^2$, where $C$ is a constant, one can observe the boundaries of Fig. \ref{ran11}(a) in system (\ref{n}) itself.  Fig. \ref{ran11}(b) confirms the existence of multi-chimera death state of oscillators in region-II.  On the other hand in the strong coupling region-III (incoherent oscillation death) phases of the oscillators are randomly distributed which implies the impact of nonisochronicity parameter on the steady states and is illustrated in Fig. \ref{ran11}(c) which resembles the PCD states observed in system (\ref{n}) (where the amplitude deviation is appreciable).  When $P= N/2$ the nonlocal coupling approaches the global limit and the corresponding phase equations are specified by \\
\begin{eqnarray}
\dot{\theta_j}=\omega -\tan \alpha+\frac{\varepsilon \tan \alpha }{2}+\frac{\xi}{2} \sin(2\theta_j-\alpha)\nonumber\\
+\frac{ \xi}{2N}\sum_{k=1}^{N}[(\sin(\theta_k-\theta_j-\alpha)-\sin(\theta_k+\theta_j+\alpha))].
\label{globaleq}
\end{eqnarray}
The above equation can be written in a more convenient form by replacing the coupling function with order parameter $R e^{i\psi}=\frac{1}{N}\sum_{j=1}^{N} e^{i\theta_j}$, where $\psi(t)$ is the phase of the global order parameter and Eq. (\ref{globaleq}) becomes 
\begin{eqnarray}
\dot{\theta_j}=\omega-c+\frac{c\varepsilon}{2}+\frac{\xi}{2} \sin(2 \theta_j-\alpha)\nonumber \\
+\frac{\xi}{2}\{ R(\sin(\psi-\theta_j-\alpha)-\sin(\psi+\theta_j+\alpha))\}.
\label{t} 
\end{eqnarray}
$\theta_j$ is the phase of the individual $j$th oscillator.  
\begin{figure*}[ht!]
\begin{center}
\includegraphics[width=0.8\linewidth]{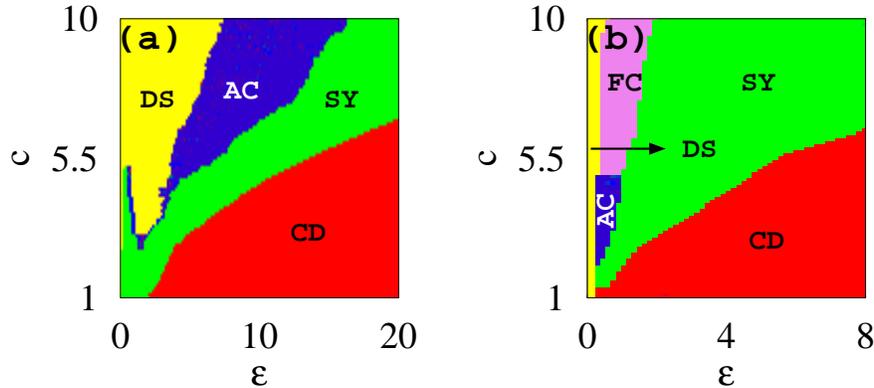}
\end{center}
\caption{(Color online) Phase diagrams of the system (\ref{n}) for (a) $r=0.1$, (b) $r=0.4$. (Hence $\frac{\varepsilon}{2P}$ varies between $0$ and $0.2$ in Fig. \ref{fg5}(a) and $0$ and $0.02$ in Fig. \ref{fg5}(b)).  SY (Green color), DS (yellow), AC (blue), FC (violet), CD (red) regions represent synchronized state, desynchronized state, amplitude chimera state, frequency chimera state, and chimera death state (CD), respectively.}
\label{fg5}
\end{figure*}
We analyze the boundary between the oscillatory and steady states by solving Eq. (\ref{t}) numerically which is plotted in the Fig. \ref{curve1}.  The solid lines with ($\vartriangle$), ($\blacksquare$) and ($\circ$) represent the numerically obtained boundaries with $N=500$, $1000$, $5000$, respectively.  As the coupling range approaches the global limit, the amplitude variations and occurrence of different states in the steady state region have disappeared.  There exists only multi-chimera death state.  By strengthening the coupling interaction, we cannot observe any appreciable change in the distribution of the multi-chimera death states as we have observed in the case of nonlocal coupling.  Thus the distribution of chimera death states does not change under strengthening of the coupling interaction under global coupling which also confirms that the occurrence of different kinds of chimera death states happens only under nonlocal coupling as shown in Sec. II.
\begin{figure}
\begin{center}
\includegraphics[width=1.0\linewidth]{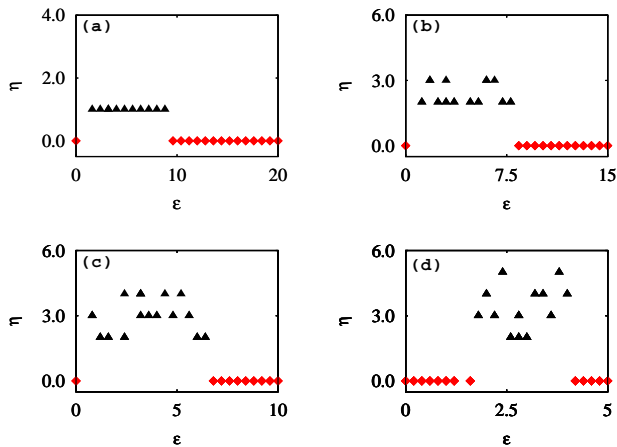}
\end{center}
\caption{(Color online) Discontinuity measure $\eta$ as a function of $\varepsilon$ with $c=5.0$ for (a) $r=0.4$, (b) $r=0.3$, (c) $r=0.2$, (d) $r=0.1$.}
\label{comb}
\end{figure} 
\begin{figure*}[ht!]
\begin{center}
\includegraphics[width=1.0\linewidth]{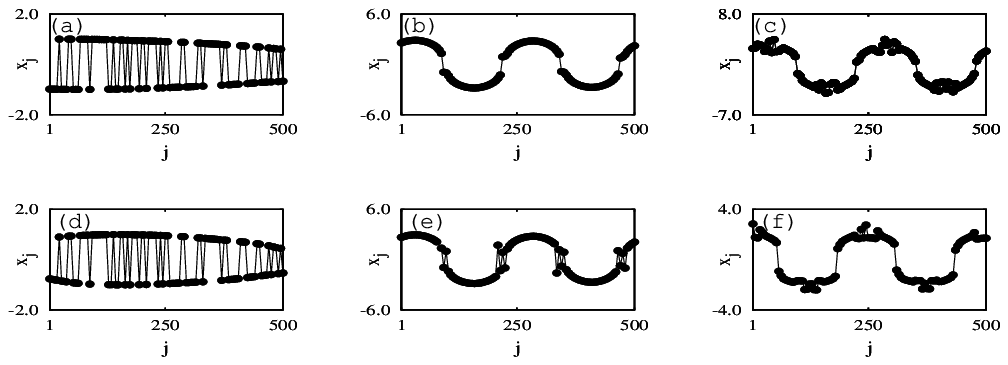}
\end{center}
\caption{The robustness of the chimera death states for $r=0.4$, $c=4$.  Distribution of frequencies of the oscillators is between $-5 \leq \omega_j \leq +5$ for (a) $\varepsilon=22.4$ ($\frac{\varepsilon}{2P}=0.056$), (b) $\varepsilon=23.2$ ($\frac{\varepsilon}{2P}=0.058$), and (c) $\varepsilon=24$ ($\frac{\varepsilon}{2P}=0.06$) and the distribution is between $-10 \leq \omega_j \leq +10$ for (d) $\varepsilon=22.4$ ($\frac{\varepsilon}{2P}=0.056$), (e) $\varepsilon=23.2$ ($\frac{\varepsilon}{2P}=0.058$), and (f) $\varepsilon=24$ ($\frac{\varepsilon}{2P}=0.06$).}
\label{fg6i}
\end{figure*} 
\section{DYNAMICS OF THE OSCILLATORY STATES}
\par In the previous section, we have studied different kinds of chimera death states under the influence of the nonisochronicity parameter with respect to the coupling range.  In addition, to study the dynamics and impact of the nonisochronicity parameter in the oscillatory region also as a function of the coupling range clearly, we first present the phase diagram in the parametric space ($\varepsilon,c$) with the help of the strength of incoherence \cite{38} $S$ (see appendix-B for more details) for $r=0.1$ in Fig. \ref{fg5}(a).  It shows that the system for finite values of $c$ ($c \leq 2.7$) is found to be synchronized by the increase of $\varepsilon$ and the symmetry breaking present in the system causes chimera death for larger $\varepsilon$.  An increase in $c$ (to $c > 2.7$) causes the  synchronized state (that appears through the increase of $\varepsilon$) to be destabilized for increase of $\varepsilon$ giving rise to  amplitude chimera state (the fluctuations exist only in the amplitudes while the frequencies of all the oscillators are the same).  Further increase of $\varepsilon$ restabilizes the synchronized state.  When the coupling strength is further increased oscillators approach the chimera death state.  For initial conditions near the synchronized state, the system attains a synchronized state.  On the other hand for initial conditions away from the synchronized state, the system attains either a chimera state or a multi-chimera death state or a periodic chimera death state by varying the coupling strength.  However, the formation of the multi-chimera death state depends on the form of the initial condition while different forms of initial conditions do not affect the formation of the periodic chimera death state.  On increasing the value of the nonisochronictiy parameter to the range $3.0<c<5.0$ one finds a suppression of the amount of spatial coherence in the system for lower values of $\varepsilon$ which leads to the onset of amplitude chimera states preceded by desynchronized states.  This indicates that strengthening of the nonisochronicity parameter causes an increase of disorder in the dynamical state.  The dynamical regions belonging to the desynchronized and chimera states are widened while strengthening the nonisochronicity parameter beyond $c=5$.
\par To know the role of $c$ for larger coupling range, we plotted the two phase diagram of the system in the ($\varepsilon,c$) parametric space for the coupling range $r=0.4$ with the help of strength of incoherence \cite{38} $S$.  Here we can find that for smaller values of $c$ the system shows direct transition from a desynchronized state to chimera death, whereas an increase in $c$ in the region $3 \leq c \leq 4.7$ causes the amplitude chimera state to intersperse this transition.  Interestingly in this case, we can observe that increasing the nonisochronicity parameter increases disorder in the frequencies of the oscillators.  Above the value of $c=4.8$, a variation in the coupling strength causes the system to transit from a desynchronized state to a synchronized state via frequency chimera state (frequency of the oscillators in the coherent regions are the same while the frequency of the oscillators belonging to incoherent regions are different) instead of amplitude chimera due to the increase of the nonisochronicity parameter which induces the disorder in the frequencies of the oscillators.  
\par  We also analyze the question whether one can observe the multiple coherent/incoherent domains (multi-chimera states) in the oscillatory region as in case of the steady state region with respect to coupling range.  To illustrate this, we make use of the discontinuity measure $\eta$ \cite{38} that will help us to distinguish the chimera and multi-chimera states.  In Fig. \ref{comb} we have plotted the discontinuity measure $\eta$ as a function of the coupling strength for different values of the coupling range ($r$).   For $r=0.4$, there occurs a coexistence of single coherent/incoherent domain in the region $0.06<\varepsilon<9.6$ ($0.0001<\frac{\varepsilon}{2P}< 0.024$), where $\eta$ takes the value one and it implies the presence of chimera state in Fig. \ref{comb}(a). There is no presence of any multi-chimera state.  Interestingly, on further decreasing the coupling range to $r=0.3$, one finds an increased number of coherent and incoherent domains in the region $0.06<\varepsilon< 8.8$ ($0.0002<\frac{\varepsilon}{2P}< 0.029$).  This indicates the existence of multi-chimera state with $\eta$ taking positive integer values greater than one.  On further decreasing the coupling range to $r=0.2$ and $0.1$, one can find several multi-coherent/incoherent domains (multi-chimera states) which are shown in Figs. \ref{comb}(c-d), respectively.  Thus we conclude that we can observe the existence of multi-chimera states depending upon the coupling range in the case of nonlocal interaction with symmetry breaking form described by (\ref{n}).
\section{STUDY OF CHIMERA DEATH IN NONIDENTICAL OSCILLATORS}
\par In order to study the robustness of the chimera death states discussed in the earlier sections, we now investigate the effect due to the presence of a perturbation in the frequency of the oscillators.  For this purpose, we consider a system of nonlocally coupled nonidentical oscillators specified by the equation   
\begin{equation}
\dot{z_j}=(1+i \omega_j)z_j-(1- ic)|z_j|^2 z_j+\frac{\varepsilon}{2P} \sum_{k=j-P}^{j+P} (Re[z_k]-Re [z_j]),
\label{nn}
\end{equation}
where $j=1,2,3...N$ and $\omega_j$ is the natural frequency of the $j^{th}$ oscillator.  We consider the case where the frequencies are uniformly distributed in a given interval.
\par  Here we address the question whether a perturbation in the frequencies of the oscillators affect the distribution of chimera death states.  To explore this point we demonstrate the case of coupling range $r=0.4$ as an example for two different perturbations.  In Figs. \ref{fg6i} (a-c), we have presented the different types of chimera death states for the frequency distribution in the range $-5 \leq \omega_j \leq +5$.  From Fig. \ref{fg6i}(a), we can observe that the nature of the multi-chimera death states does not change qualitatively.  But there occurs an increase of the nonuniformity in the distribution of the inhomogeneous steady states in the weak coupling limit. One can also note that there is not much qualitative change in the type-I PCD and type-II PCD regions which is clearly shown in Figs. \ref{fg6i}(b,c). 
\par  For the choice of the frequency distribution between  $-10 \leq \omega_j \leq +10$, again there is an increase of disorder in the distribution of multi-chimera death region [Fig. \ref{fg6i} (d)].  We can also note that there is no change in the formation of both type-I and type-II periodic chimera death states which are clearly seen in Figs. \ref{fg6i} (e, f).  Thus we conclude that perturbations in the natural frequencies of the oscillators do not affect the nature of the chimera death states in the strong coupling limit which indicates that the inhomogeneous steady states are stable to such perturbations.  On the other hand, in the case of multi-chimera death states, there occurs an increase in the nonuniformity to even small perturbations in the natural frequencies of the oscillators.
\begin{figure*}[ht!]
\begin{center}
\includegraphics[width=1.0\linewidth]{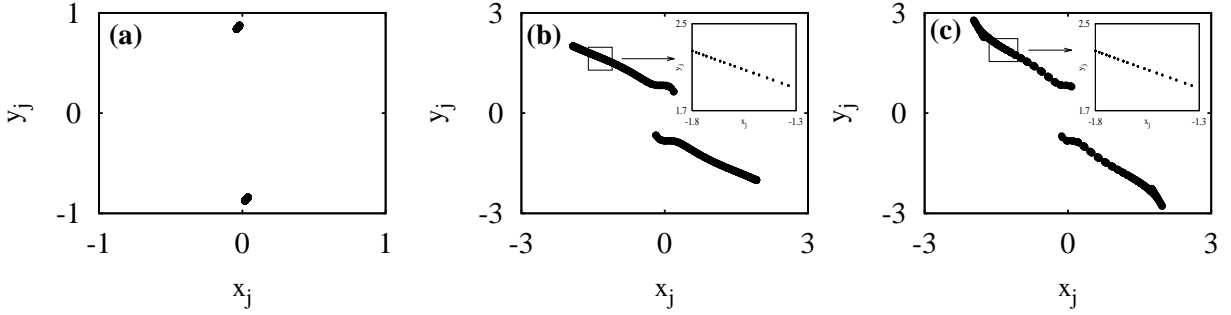}\\
\end{center}
\vspace{-0.6cm}
\caption{Phase portrait of the oscillators in the ($x,y$) plane with $c=4$ and $r=0.4$: (a) multi-chimera death state for $\varepsilon=22.4$ ($\frac{\varepsilon}{2P}=0.056$), (b) type-I periodic chimera death state for $\varepsilon=23.2$ ($\frac{\varepsilon}{2P}=0.058$) and (c) type-II periodic chimera death state for $\varepsilon=24$ ($\frac{\varepsilon}{2P}=0.06$).}
\label{fgt2}
\end{figure*} 
\section{conclusion}
\par In summary, we have investigated the occurrence of different types of chimera death states in a nonlocally coupled network of Stuart-Landau oscillators with symmetry breaking form.  We have analyzed the impact of the nonisochronicity parameter on the chimera death states which brings out the amplitude and phase variations and results in multi-chimera death states (coupling strength less than the critical value), type-I PCD states and,  type-II PCD states in the strong coupling limit (coupling strength greater than the critical value).  In addition, we have also identified that the number of periodic domains in both the PCD states follow a power law relation with the coupling range only under nonlocal coupling rather than local and global couplings while multi-chimera states which occur in the oscillatory region do not follow any such relation.  Moreover, we have also shown that the perturbations in the natural frequencies of the oscillators do not affect the nature of the chimera death states in the strong coupling limit while it increases the inhomogeneity in the distribution of chimera death states in the weak coupling limit. \\
\section*{Acknowledgements}
The work of  KP and MS forms part of a research project sponsored by NBHM, Government of India.  The work of VKC is supported by the SERB-DST Fast Track scheme for young scientists under Grant No. YSS/2014/000175.  ML acknowledges the financial support under a DAE Raja Ramanna Fellowship program.
\appendix
\section{ANALYSIS OF THE AMPLITUDE VARIATIONS IN THE CHIMERA DEATH REGION}
\par We can also confirm the existence of amplitude variations caused by the nonisochronicity parameter by plotting the phase portrait of the oscillators in the ($x_j,y_j$) plane which is illustrated in Fig. \ref{fgt2}.  We can observe from Fig. \ref{fgt2}(a) that the distribution of inhomogeneous steady states have same value for multi-chimera death state.  On the other hand, there occurs variations in the distribution of inhomogeneous steady states as shown in Figs. \ref{fgt2}(b, c) for type-I and type-II periodic chimera death states, respectively. 
\section{CHARACTERISTIC MEASURE: STRENGTH OF INCOHERENCE}
\par In order to know the nature of dynamical states in more detail, we look at the strength of incoherence of the system a notion introduced recently by Gopal, Venkatesan and two of the present authors\cite{38}, that will help us to detect interesting collective dynamical states such as synchronized state, desynchronized state, and the chimera state.  For this purpose we introduce a transformation $z_j=x_j-x_{j+1}$ \cite{38}, where $j=1,2,3,...,N$.  We divide the oscillators into $M$ bins of equal length $n=N/M$ and the local standard deviation $\sigma(m)$ is defined as   
\begin{equation} 
\sigma(m)=\langle(\overline{ \frac{1}{n}\sum_{j=n(m-1)+1}^{mn} \vert z_j-\overline{z_j}\vert^2})^{1/2}\rangle_t, m=1,2,...M.
\label{sig}
\end{equation}
\par From this we can find the local standard deviation for every $M$ bins of oscillators that helps to find the strength of incoherence \cite{38} through
\begin{equation} 
S=1-\frac{\sum_{m=1}^{M} s_m}{M},s_m=\Theta(\delta- \sigma(m)),
\label{soi}
\end{equation}
where $\delta$ is the threshold value which is small. When $\sigma(m)$ is less than $\delta$, $s_m$ takes the value $1$, otherwise it is $0$. Thus the strength of incoherence measures the amount of spatial incoherence present in the system which is zero for the spatially coherent  synchronized state.  It has the maximum value, that is $S=1$, for the completely incoherent desynchronized state and has intermediate values between 0 and 1 for chimera states and cluster states.  These concepts have been well verified in the case of many dynamical systems, see for example \cite{38}.  In our present study we have taken the number of bins $M=100$ and $\delta=0.05$.
\par To distinguish further between chimera and multi-chimera states, we also use a discontinuity measure, based on the distribution of $s_m$ in (\ref{soi}). It is defined as
\begin{equation}
\eta=\frac{\sum^M_{i=1}|s_i - s_{i+1}|}{2}, (s_{M+1}=s_1).
\label{eta}
\end{equation}
 In this case $\eta$ takes a value zero for a coherent or incoherent state, unity for chimera state, and positive integer values greater than one $(2 \leq \eta \leq M/2)$ for multi-chimera states (due to the fact that how many chimera regions exist for a given value of coupling parameter so that the difference between $s_i$ and $s_{i+1}$ at the beginning and end of each region contribute a factor $+1$ to Eq. (\ref{eta})).

\end{document}